# ELECTROCHEMICAL BEHAVIOUR OF MONOLAYER AND BILAYER GRAPHENE

Anna. T. Valota[†], Ian A. Kinloch[‡], Kostya S. Novoselov[§], Cinzia Casiraghi[†], Axel Eckmann[†], Ernie W. Hill[∥] and Robert A.W. Dryfe[†]*

† = School of Chemistry

‡ = School of Materials

§ = School of Physics & Astronomy

∥ = School of Computer Science

University of Manchester, Oxford Road, Manchester M13 9PL,

United Kingdom.

e-mail: robert.dryfe@manchester.ac.uk

Fax: +44 161 275 4598

**Abstract:** Results of a study on the electrochemical properties of exfoliated single and multilayer graphene flakes are presented. Graphene flakes were deposited on silicon/silicon oxide wafers to enable fast and accurate characterization by optical microscopy and Raman spectroscopy. Conductive silver paint and silver wires were used to fabricate contacts; epoxy resin was employed as masking coating in order to expose a stable, well defined area of graphene. Both multilayer and monolayer graphene microelectrodes showed quasi-reversible behavior during voltammetric measurements in potassium ferricyanide. However, the standard heterogeneous charge transfer rate constant, k°, was estimated to be higher for mono-layer graphene flakes.

**Keywords:** Graphene, electrochemistry, Raman spectroscopy, electron transfer.

The unique charge transport properties of graphene have attracted enormous interest. Amongst the many conceived applications of graphene, there is also much interest in its use as an electrode material, with energy storage/conversion applications in mind. Examples include recent reports of graphene-based supercapacitors[1-3] and of the use of graphene as (almost) transparent electrodes in solar cells.[4-9] Optimisation of such applications requires complete understanding of the electron transfer properties of graphene. However, despite a considerable amount of effort applied in this direction,[10-15] the understanding of the electrochemical activity of graphene remains controversial.[16] This work in fact feeds in to the considerable debate about the relative activity of the basal plane and edge planes of (bulk) graphitic samples.[17-21] Much of the lack of clarity about the electrochemical properties of graphene stems from the rather ill-defined sample preparation of almost all of the earlier reports on graphene electrochemistry: generally, graphene flakes (often a mixture of monolayer, bilayer and "other" samples) are simply dispersed on a conducting substrate, with no attempt made to isolate individual flakes, or to isolate the flake edges from the basal plane. Furthermore, the majority of studies to date have obtained graphene *via* chemical means, usually by reduction of graphene oxide, leaving doubts about the purity of the material.[22-24] There are a few notable exceptions to this general trend. Work on the capacitance of graphene has used a "top-gated" (effectively, electrochemical) configuration to bias the graphene/electrolyte interface.[25, 26] More recently, work has appeared which describes the electrochemical properties of a single, masked flake in contact with an aqueous solution of a ferrocene derivative.[27] In the latter article, the authors noted an increase in electron transfer rate for the oxidation of the ferrocene derivative on the graphene sample, compared to values quoted on "bulk" graphite (highly oriented pyrolytic graphite, HOPG). The authors were unable to measure a rate constant for electron transfer on mechanically exfoliated graphene, the process was sufficiently fast that it was essentially reversible. A finite (slower) rate constant was determined on graphene prepared by chemical vapour deposition methods, this rate was still substantially higher than that measured on basal plane HOPG. The enhanced electron transfer kinetics seen on graphene were ascribed to the ripples present on its surface by the authors of this work. In this report, we have directly compared the voltammetric responses on a multilayer graphitic surface, with the response obtained for monolayer and bilayer graphene. Ferricyanide, which has been reported to show poor charge transfer kinetics on basal plane HOPG,[18,19] has been chosen as a model system to highlight differences in the behavior of the solids. Accordingly the first report of charge transfer kinetics on a monolayer graphene sample are presented, confirming that electron transfer kinetics are indeed improved on the graphene surface, relative to bulk graphitic materials.

**Results:** The reduction of aqueous phase ferricyanide was selected as our model redox process. A reduction process was chosen, to avoid the risk of oxidation of the graphene samples. Furthermore, a redox couple with a relatively low standard reduction potential was chosen to minimise any interference due to the reduction of the solvent background. The ferricyanide couple is interesting because it has been extensively investigated on carbon surfaces and is reported to have slow electron transfer kinetics on the basal plane of well-defined bulk graphitic phases (specifically, highly oriented pyrolytic graphite, HOPG).[19] A schematic diagram showing an overview of the analyzed samples is presented in Figure 1. Samples may be classified as defect-free monolayers, with all the edges covered by the masking resin (Figure 1(a, b)), defective monolayers with evident holes (Figure 1(c, d)), monolayers with exposed edges (Figure 1(e, f) and multilayers, which for the purposes of this diagram means two or more layers (Figure 1(g, h)). Optical micrographs of a multilayer sample (> 20 graphene sheets) are shown in Figure 2. Flakes of natural graphite were produced with a thickness varying from tens to hundreds of nanometers. AFM topography has previously revealed that steps and folds generated by cleavage of natural graphite are usually 10-20 nm high, with about one step edge every 800 nm.[28] Such defects are evident on the multilayer sample in Figure 2(a), before exposure to the solution. Special attention was paid during masking of samples in order to expose areas with the minimum number of defects, however to date it has not been possible to achieve a perfect, edge-free region. However, Figure 2(b) indicates that no obvious change in the sample was observed following voltammetry. Raman spectra of the sample are presented in Figure 3. The voltammetric response observed on our multilayer sample is entirely consistent with previous literature reports for the ferri/ferrocyanide couple on (bulk) graphite and will be considered as our reference (see Figure 4(a)).[18,19] The limiting, background-subtracted current obtained at a scan rate of 5 mV s$^{-1}$ was 7.5 nA. The predicted diffusion-limited current for an inlaid microdisc electrode varies between 6.2 and 9.5 nA, according to equation [1]:[29]

$$I_{lim} = 4nFDcr \qquad [1]$$

Where *n* is the number of electrons exchanged in the redox reaction (1 in the present case), *F* is the Faraday constant, *D* is the diffusion coefficient of the electroactive species in solution (reported to vary between $5.37 \times 10^{-6}$ cm$^2$ s$^{-1}$ [30] and $8.20 \times 10^{-6}$ cm$^2$ s$^{-1}$ [31] for this solute), *c* is the bulk concentration of the ferricyanide and *r* is the radius of the window to the solution. The standard electron transfer rate constant, $k_0$, for ferricyanide reduction on the multilayer graphite surface was calculated, from the procedure described by Mirkin and Bard,[32] to be 7 $\times 10^{-4}$ cm s$^{-1}$. Ferricyanide reduction on graphite has been widely studied in literature, most notably by McCreery and co-workers.[19,33] For HOPG with very low defect density, standard electron transfer rate constants lower than $10^{-6}$ cm s$^{-1}$ were found. However, a 1% defect density is estimated to cause a $10^3$ factor increase in $k_0$.[18]

The voltammetric responses of various monolayer graphene samples, which had different levels of defect visible, were investigated using the same redox couple (ferri/ferrocyanide, also shown in Figure 4). Micrographs of these samples are shown in Figure 5. Monolayer sample 1, shown in Figure 5(a) and (b) contained no visible defects and its edges were completely masked, a conclusion supported by the fact that the ratio between the intensity of D and G peaks was lower than 0.1 in each point of the Raman map[34] (see Figure 6(d)). Before the masking process (Figure 5(a)), the surface of graphene Sample 1 appeared as homogeneous, with a characteristic colour.[35] After masking was completed, the presence of bright dots was revealed by optical microscopy (Figure 5(b)). The nature of the observed dots, transparent to Raman spectroscopy (see Figure 6(a) and (b)) is unclear. However, the appearance of the dots just after the masking process indicates possible contamination of the graphene surface by epoxy particles.

Figure 7(a) shows the background voltammetric response of Sample 1 (masked, defect-free). Measurement of the non-Faradaic current density (at 0 V) as a function of sweep rate gives an estimate of the total interfacial capacitance per unit area, found to be 21.3 µF cm$^{-2}$ (Figure 7(b)). In aqueous electrolytes, the total interfacial capacitance is composed of the quantum capacitance contribution from the graphene in series with the capacitance of the solution double layer, which can be resolved into a diffuse and a compact component.[36] In the case of concentrated electrolytes such as the one used in the present work (3 M), the capacitance of the diffuse ionic layer is usually large (>100 µF cm$^{-2}$),[36] therefore its contribution to the total capacitance is negligible. The capacitance arising from the compact layer is known to have a value of about 10-20 µF cm$^{-2}$,[36] suggesting that the value quoted above for the total interfacial capacitance is anomalously high. We note that capacitance of graphene has previously been quoted as between 8 and 10 µF cm$^{-2}$ in 1 mM NaF aqueous solution, employing a. c. impedance spectroscopy, hence further investigation is required in order to clarify the aqueous phase capacitance behaviour of monolayer graphene.[25]

Ferricyanide voltammetry at Sample 1 is shown in Figure 4(c). As with the multilayer sample, the limiting current here (8.5 × 10$^{-9}$ A), calculated *via* equation [1], is in reasonable agreement with the experimental value (9.6 × 10$^{-9}$ A), assuming a ferricyanide diffusion coefficient of 5.4 × 10$^{-6}$ cm$^2$ s$^{-1}$.[30] However, the current-potential response is more reversible than the multilayer graphite case, see Figure 4(b) and (d), indicating that the graphene surface – in spite of its apparent lower level of defects – acts as an efficient catalyst for electron transfer from ferricyanide. A $k_0$ value of 1.2 × 10$^{-3}$ cm s$^{-1}$ was found for the monolayer sample (Sample 1), almost twice as high as the standard rate of electron transfer estimated with the (defect containing) multilayer.

As a matter of control we have also prepared samples with a number of defects. For instance, monolayer Sample 2 presented in Figure 5(c) contains several holes of ca. 10 micron diameter, hence some edge sites must be in contact with the electrolyte; the exposed part of monolayer Sample 3 is triangular (Figure 5(d)), hence edges are also exposed to solution in this case. Surprisingly, in view of received wisdom about the role of defects on electron transfer rates for bulk graphitic samples, the voltammetric responses of the defective monolayer samples (Samples 2 and 3) were not markedly different from that of the defect-free sample (Sample 1), at least for the case of ferricyanide reduction. Figure 8 presents the comparative current-potential response, where the current is normalised by sample radius to account for the different exposed windows of the monolayer samples (c.f. equation [1]). The current data obtained from a bilayer graphene sample is also presented, indicating that this material also presents electron transfer kinetics which are more similar to monolayer graphene than to the multilayer graphitic material.

It should be noted that the electron transfer kinetics obtained at the monolayer graphene samples degrade somewhat over a period of time. Figure 8 also shows the response from the monolayer Sample 2 (few holes, see Fig. 5(c)) after a two week exposure to ambient conditions. The normalised current response is almost identical to that seen for the multilayer sample. A further feature we have observed is that defects on the graphene surface appear to act as nucleation sites for the fracture of samples under the influence of the applied potential. This phenomenon is obviously of interest and is the focus of ongoing work.

**Conclusions:**

Monolayer and bilayer samples of graphene are electroactive and present improved electron transfer kinetics, for the case of ferricyanide reduction, compared to the basal plane graphite substrates. Defects present on the monolayer make little difference to the voltammetric response of the samples.

**Acknowledgements:** We thank the U.K. EPSRC (grant references EP/I005145/1 and EP/G035954/1), the Royal Society and the University of Manchester EPS strategic equipment fund for financial support. We thank J. Martin for assistance with preliminary experiments.

**Experimental Methods:**

*Fabrication of Electrodes:* Samples of monolayer and bilayer graphene, and multilayer (> 20 graphene sheets) graphite, were prepared by mechanical exfoliation of natural graphite supplied by NGS Naturgraphit GmbH. The samples were transferred to a silicon wafer, covered with a 90 nm thick thermal oxide layer. Conductive silver paint and silver wires were used to fabricate contacts. The samples were masked with an epoxy resin to leave a window of the order of 50 microns in diameter, deliberately exposing either the basal plane of each sample, or the basal plane and some of its edges. The precise dimensions of each exposed window were determined by optical microscopy. The samples were characterised by Raman spectroscopy, either using a Renishaw spectrometer (50× objective, ~0.7 mW power) or a Witec spectrometer (100× objective, ~0.6 mW power) at an excitation wavelength of 633 nm.

*Chemicals and Electrochemical Apparatus:* Voltammetric experiments were performed in aqueous solution using a three electrode configuration under potentiostatic control (Autolab

PGSTAT30, Utrecht, the Netherlands). The masked graphene/graphite samples were used as the working electrode, an Ag/AgCl wire (prepared in-house) was used as the reference electrode, a Pt gauze was employed as the counter electrode. Water was obtained from an ELGA PureLab-Ultra purifier (minimum resistance 18.2 MΩ cm). For the electrolyte solutions, the redox active salt, $K_3Fe(CN)_6$, and the supporting electrolyte, KCl, were purchased from Sigma-Aldrich and Fisher-Scientific, respectively, and used as received. The pH of the freshly prepared ferricyanide electrolyte solution was 5.8 (Hanna Instruments pH meter).

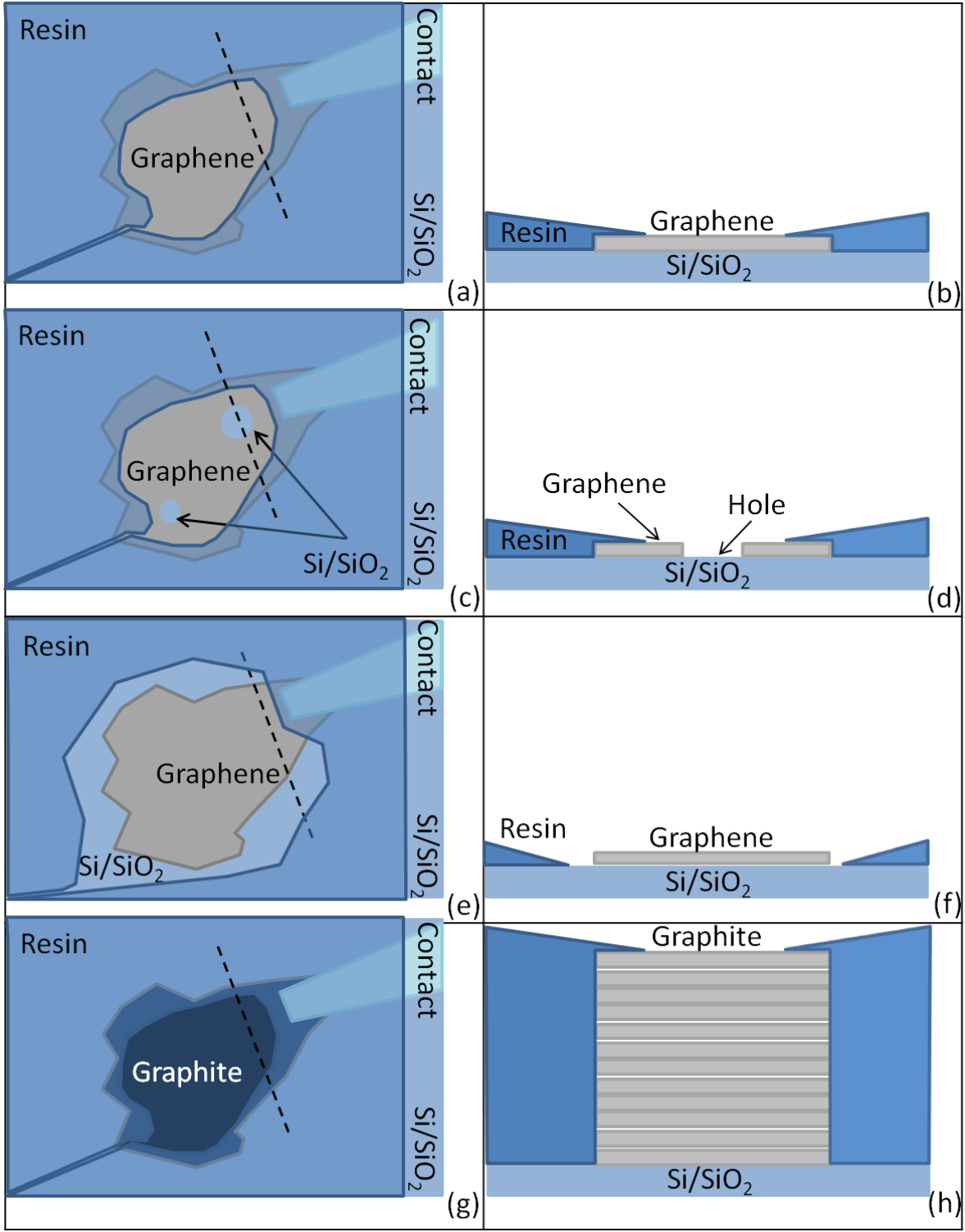

**Figure 1** - Schematic diagram of the samples employed in the present work. On the left: top view of the samples; on the right: cross section along the black dashed line. Samples are classified as: defect-free monolayers, with all the edges covered by the masking resin (a, b); defective monolayers with evident holes (c, d); monolayers with exposed edges (e, f) and multilayers (g, h).

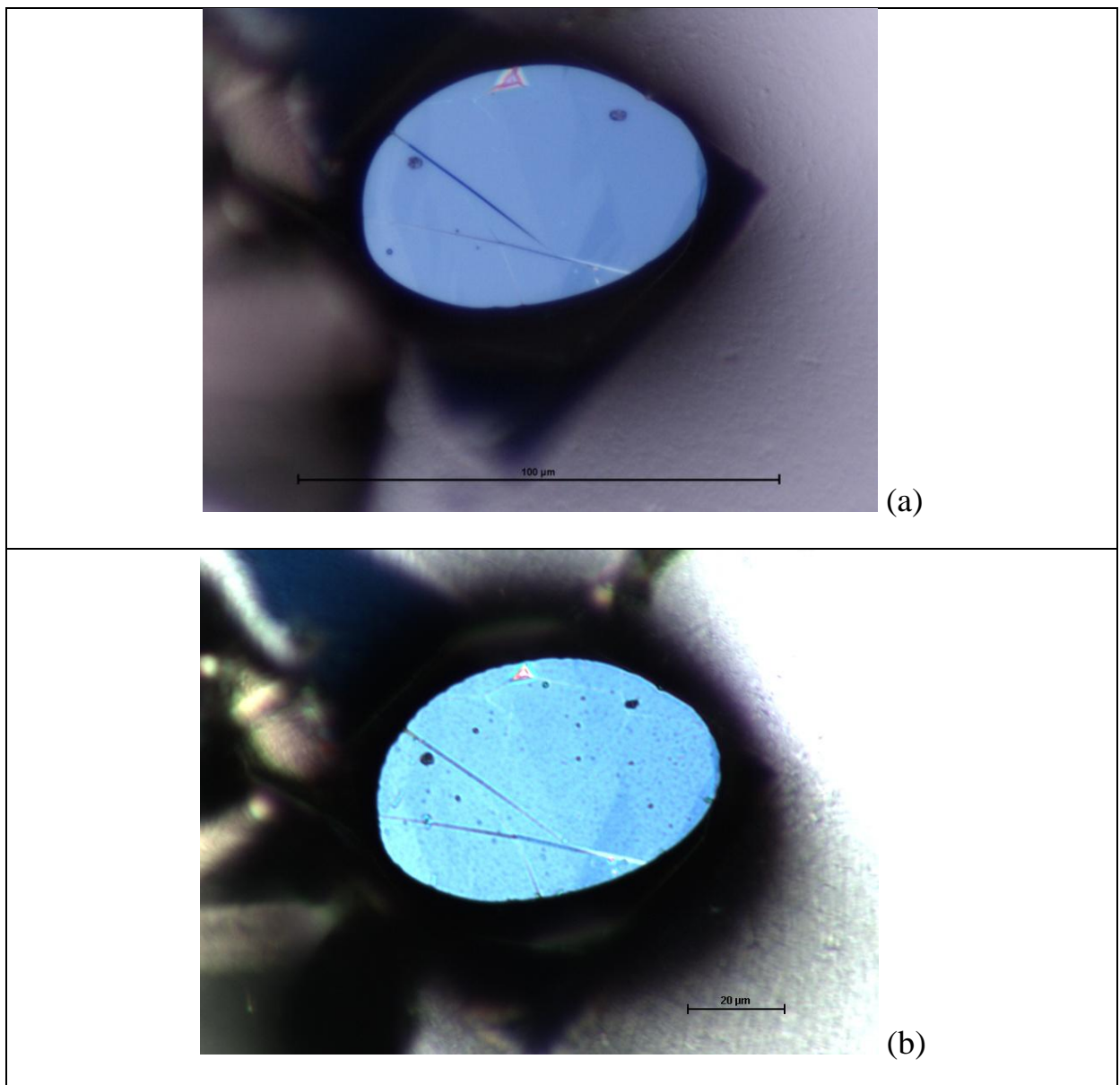

**Figure 2 -  Optical Micrographs of the multilayer sample before (a) and after (b) voltammetric experiments.**

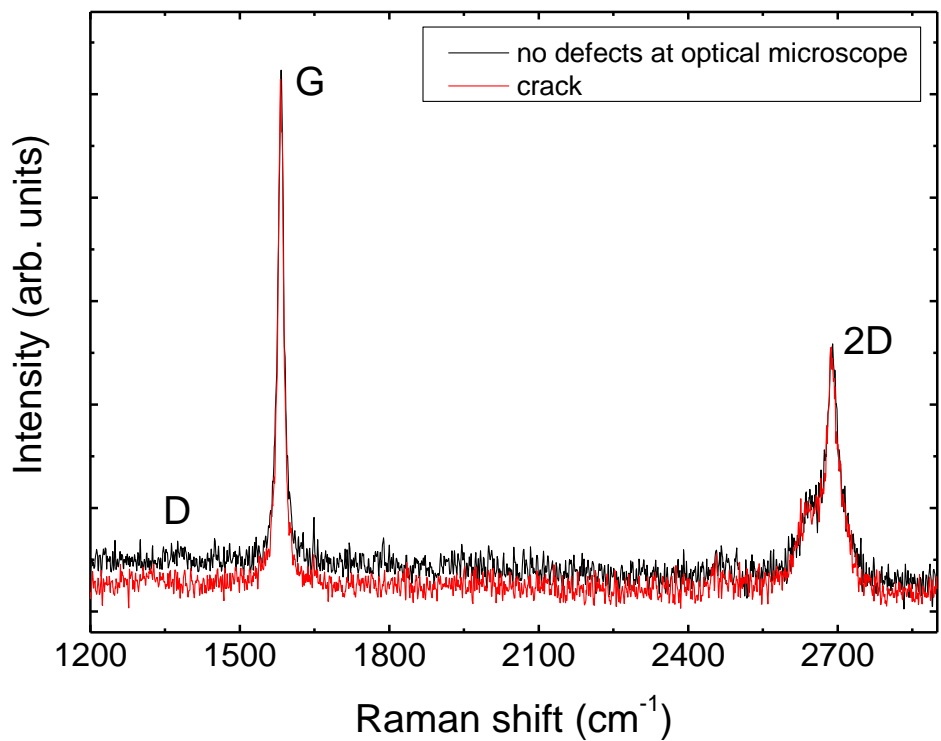

**Figure 3 - Raman spectra performed on two different spots on the multilayer graphene sample. Excitation wavelength: 633 nm, 50 × objective, ~0.7 mW power (Renishaw spectrometer).**

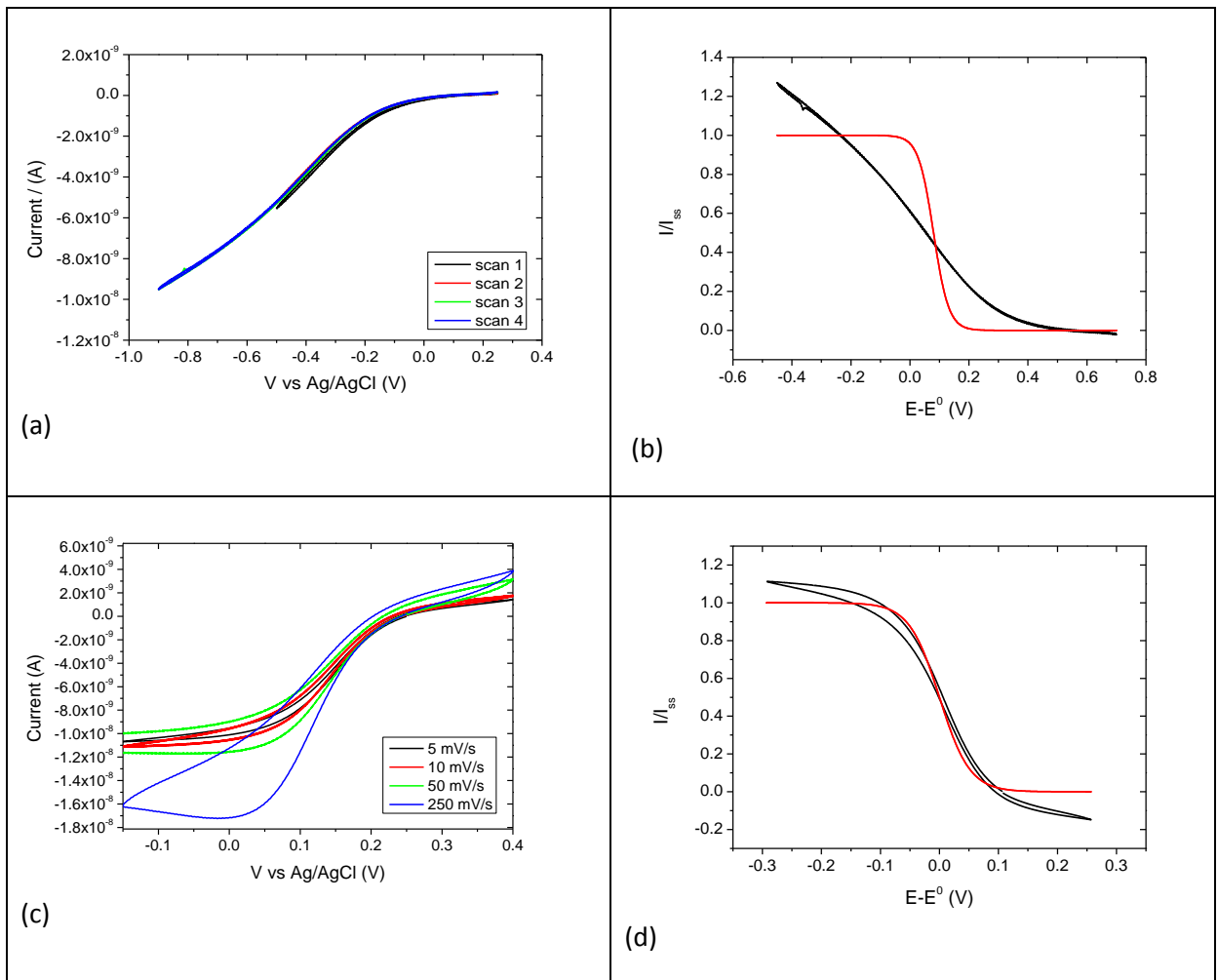

**Figure 4** - Voltammetric response of multilayer graphene at 5 mV/s (a) in 1 M KCl and 1 mM ferricyanide electrolyte; (b) the experimental data (black lines) is plotted along with the ideal response for reversible electron transfer (red lines). (c) The voltammetric data for various scan rates (see panel) is shown for the monolayer sample 1, the defect-free graphene; (d) applies the analysis performed for the multilayer sample in (b) to the 5 mV/s data from Sample 1. In both (b) and (d), currents are normalised to the steady-state current.

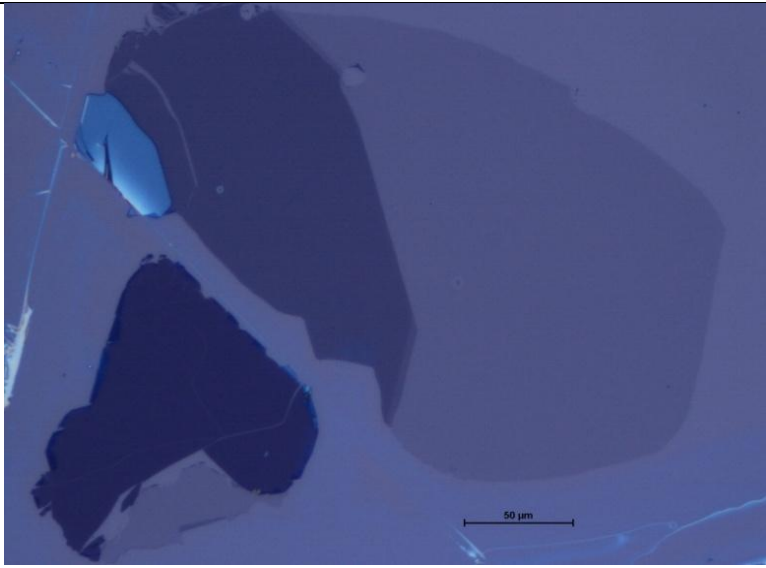

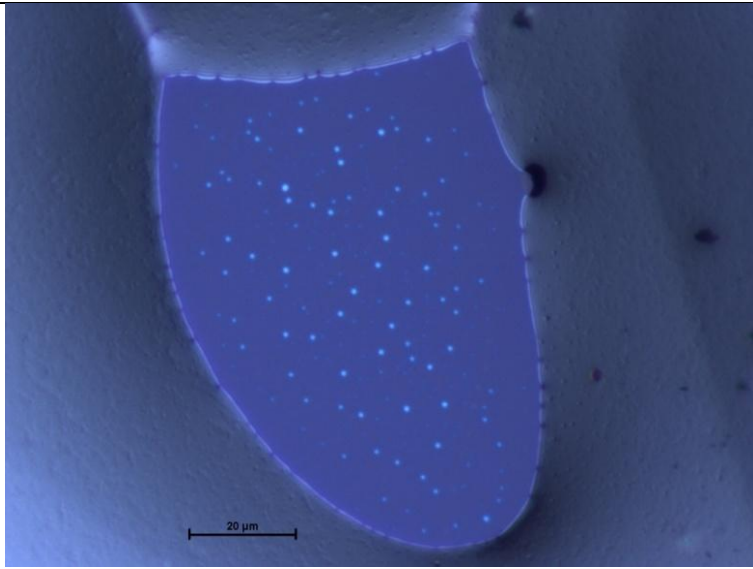

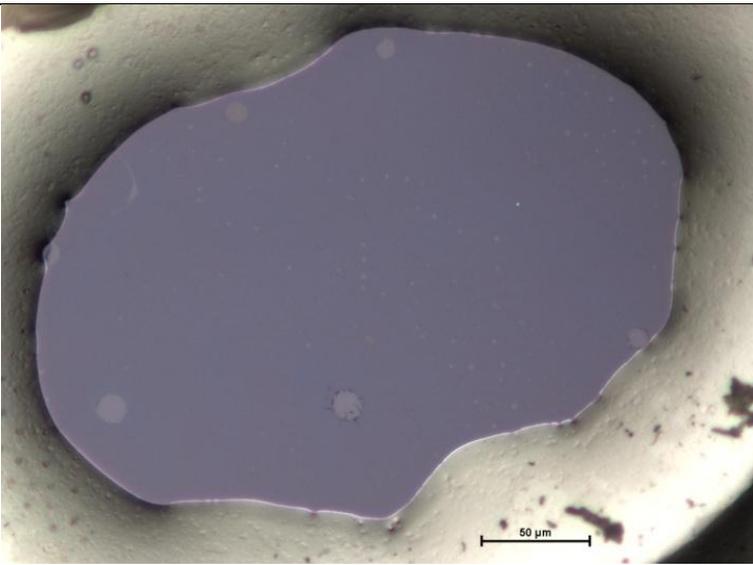

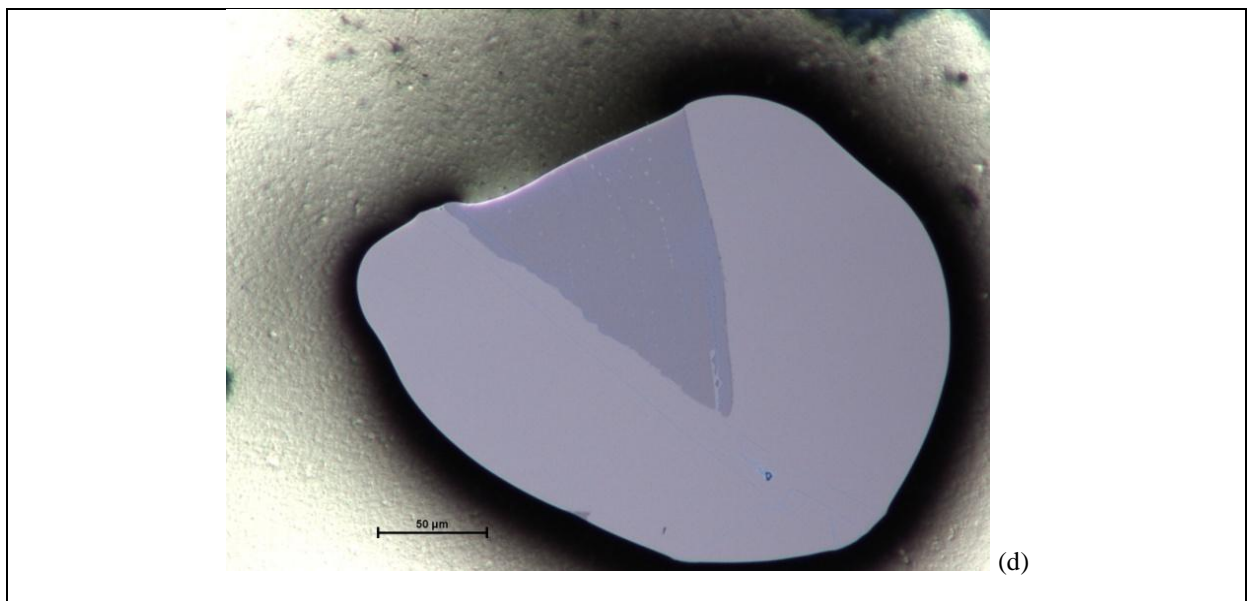
(d)

**Figure 5 - Optical micrographs of monolayer graphene samples. Sample 1 is shown before (a) and after (b) masking; in image (b) edges are completely masked. Sample 2 is shown to contain holes, in (c); the exposed part of sample 3 is triangular, hence edges are exposed to solution, see (d).**

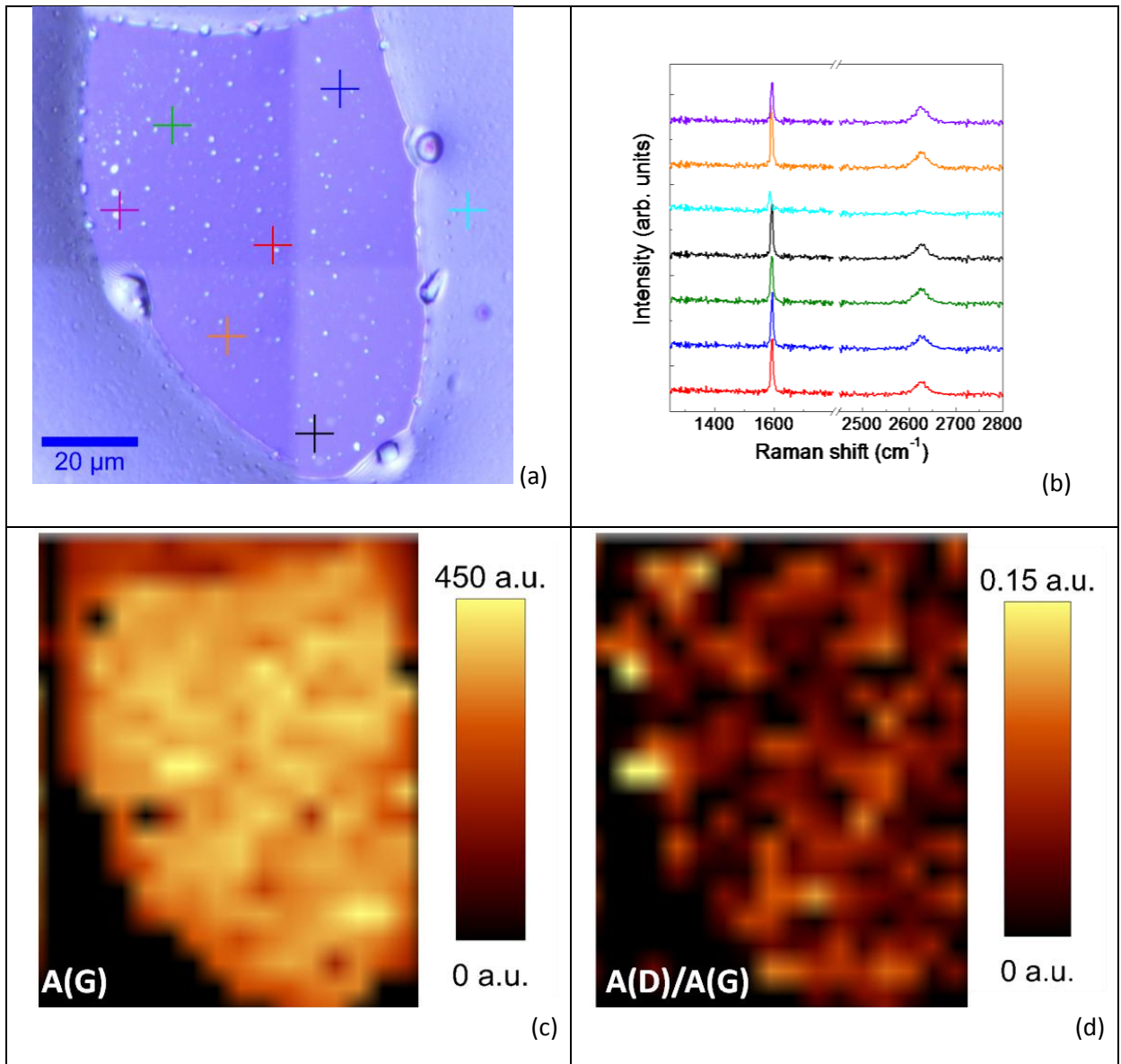

Figure 6 – Optical image (a) and Raman spectra (b) performed on seven different spots on the defect-free monolayer graphene (Sample 1), after the first series of electrochemical measurements. Excitation wavelength: 633 nm, 100 × objective, 0.7 mW laser power (Witec spectrometer). The crosses correspond to the locations where the mapping was performed; the colours of the crosses correspond to the colours of the Raman spectra. Map of the G peak integrated area (c) and of the ratio between the D peak and G peak integrated areas (d).

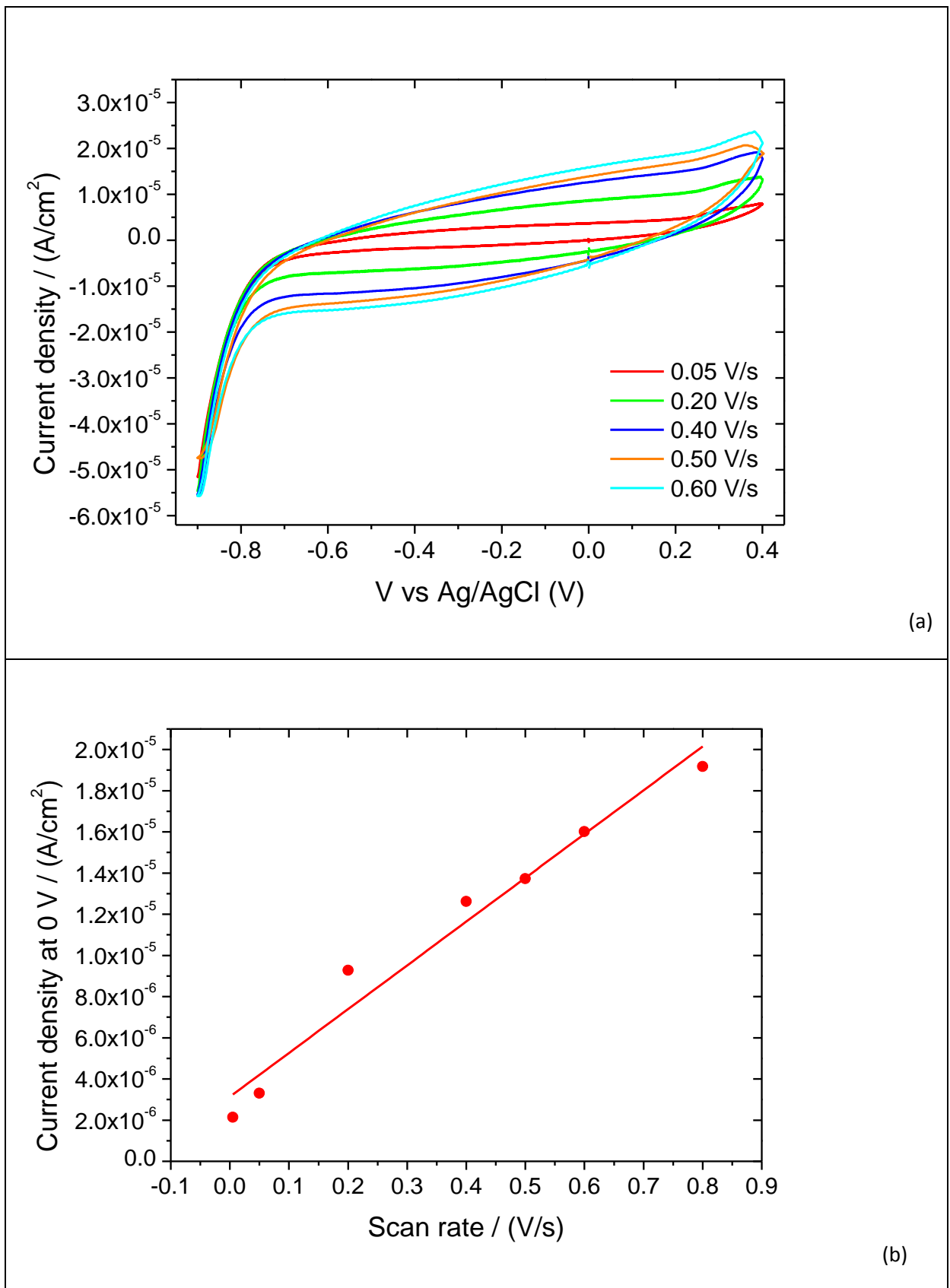

**Figure 7 -  Current response of defect-free monolayer graphene (Sample 1) in background electrolyte (3 M aqueous solution of KCl) at various scan rates (a); plot of current density measured at 0 V *versus* scan rate (b). The gradient of the fitted line indicates a capacitance of 21.3 µF/cm$^2$.**

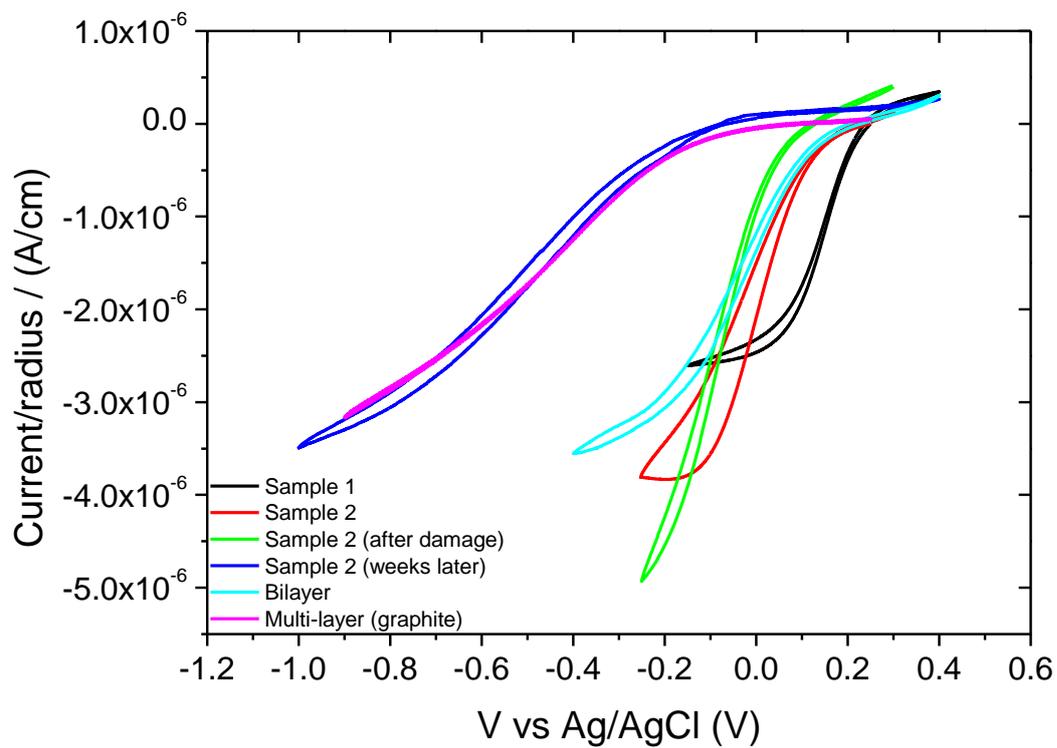

**Figure 8 - Current (normalised to electrode radius)** *vs.* **potential response for the graphene monolayer samples (Samples 1 and 2), a bilayer sample and the multilayer.**